\documentclass[aps,pop,reprint,amsmath,amssymb,floatfix]{revtex4-2}

\usepackage{graphicx}
\usepackage{dcolumn}
\usepackage{bm}
\usepackage{hyperref}
\usepackage{listings}
\usepackage{xcolor}
\usepackage{booktabs}

\begin{document}

\preprint{arXiv preprint}

\title{NeSST: A Python Tool for Neutron Spectra\\
       and Synthetic Diagnostics in Inertial Confinement Fusion}

\author{A.~J.~Crilly}
\email{ac116@ic.ac.uk}
\affiliation{Centre for Inertial Fusion Studies, The Blackett Laboratory,
             Imperial College London, London SW7~2AZ, United Kingdom}

\date{\today}

\begin{abstract}
We present the Neutron Scattered Spectra Tool (NeSST), an open-source Python package for rapidly constructing primary and singly scattered neutron spectra from inertial confinement fusion (ICF) implosions. NeSST evaluates primary spectra for deuterium--tritium (DT), deuterium--deuterium (DD) and tritium--tritium (TT) reactions. Differential and total nuclear cross sections are read directly from Evaluated Nuclear Data File (ENDF) libraries. This enables elastic ($n$D, $n$T) and inelastic [D$(n,2n)$p, T$(n,2n)$D] scattering from DT fuel, as well as scattering from additional ablator materials such as $^{12}$C, to be treated within a common framework.  Relativistic corrections to elastic scattering kinematics are included. Areal density asymmetries are incorporated through a Legendre mode expansion of the neutron-averaged projected areal density, allowing the spectral signatures of implosion non-uniformities to be computed and fitted.  The effect of scattering ion velocities on the neutron backscatter edge shape is handled through pre-computed ion-velocity-dependent scattering kernels.  A synthetic neutron time-of-flight (nToF) module converts energy spectra into detector signals with a full forward model that includes configurable instrument response functions (IRFs), energy-dependent scintillator sensitivity, and beamline attenuation. The code is publicly available at \url{https://github.com/aidancrilly/NeSST}.
\end{abstract}

\maketitle

\section{\label{sec:intro}Introduction}

Neutron spectroscopy\cite{GatuJohnson2013, forrest2012, Frenje2010, gatu2024} is a primary diagnostic for inertial confinement fusion (ICF) experiments, providing direct measurements of fuel ion temperature, areal density, bulk flow velocity, and fusion yield.  The measured spectrum contains contributions from primary (unscattered) fusion neutrons and from neutrons that have undergone one or more nuclear collisions within the fuel or surrounding ablator.

For DT implosions, the primary DT peak near 14~MeV encodes the hotspot ion temperature (through its spectral width) and the bulk flow velocity\cite{Hatarik2018,Mannion2018} (through its centroid shift).  The ratio of the down-scattered to primary neutron yield, the down-scatter ratio (DSR), measures the neutron-averaged fuel areal density $\langle\rho L\rangle$~\cite{GatuJohnson2013}.  In addition to the broad elastic scattering continuum, neutron backscatter edges from $n$T and $n$D elastic collisions appear at 3.5 and 1.56 MeV and are sensitive to the ion velocity distribution in the dense fuel shell\cite{Crilly2020,Crilly2022,Crilly_POP2024}.  The 2.45~MeV DD primary peak, from $D+D\to{}^3\mathrm{He}+n$, and the broad TT spectrum, from $T+T\to{}^4\mathrm{He}+2n$, also appear at significant levels in a DT plasma and carry complementary diagnostic information.  In the regime of moderate areal density ($\langle\rho L\rangle \lesssim 200~\text{mg/cm}^2$), the single-scatter contribution dominates over higher-order scattering and carries detailed information about fuel conditions\cite{Crilly2021,Gopalaswamy2025,Forrest2022}.

Several tools exist for computing ICF neutron spectra, ranging from full Monte Carlo neutron transport calculations (such as MCNP\cite{mcnp}, OpenMC\cite{openmc} and IRIS3D\cite{Weilacher2018}) to deterministic solvers for the neutron transport equation (such as Minotaur\cite{Crilly2018}).  These higher-fidelity tools capture multi-scatter contributions, spatial transport, and geometry effects that are beyond the scope of the NeSST package, but they can be computationally expensive, making rapid iteration over parameter space challenging.  NeSST fills the complementary niche of a lightweight, analytical/semi-analytical framework that is intentionally limited to the single-scatter regime, enabling synthetic signal generation and fitting without significant computational overhead.  Because its scattering matrices are pre-computed and subsequent spectrum evaluations reduce to matrix--vector multiplication (via trapezoid rule integration), NeSST is particularly suited to parameter-space searches, uncertainty quantification, and the generation of large libraries of synthetic diagnostic signals within the single-scatter approximation.

The physics models underlying NeSST’s scattering-matrix approach were developed in two companion papers: Crilly~et~al.~\cite{Crilly2020} developed the model for the neutron backscatter edge and its dependence on dense fuel ion kinematics, and Crilly~et~al.~\cite{Crilly2021} extended the framework to incorporate areal density asymmetries and their spectral signatures. NeSST also makes use of canonical models for the primary spectra within a DT plasma\cite{Brysk1973,Ballabio1998}. The present paper describes the implementation of these models as a publicly available software package, together with additional capabilities---notably multiple TT primary spectral models (based on Brune's R-matrix calculation~\cite{brune2015}), support for additional scattering materials (e.g.\ $^{12}$C ablator) via bundled ENDF data, an approximate second-scatter contribution for higher-$\langle\rho L\rangle$ conditions, and a full synthetic nToF diagnostic module with physically motivated IRF components---that have been added since those publications. 

The remainder of this paper is organised as follows. Section~\ref{sec:primary} describes the primary neutron spectrum models.  Section~\ref{sec:scatter} develops the single-scatter spectral model, including the areal density geometry, cross-section treatment, and asymmetry formalism. Section~\ref{sec:ionkin} describes the ion kinematics model for the backscatter edge. Section~\ref{sec:highrhoL} describes approximate models for extending NeSST's predictions to higher areal densities. Section~\ref{sec:materials} discusses support for additional scattering materials. Section~\ref{sec:ntof} presents the synthetic nToF diagnostic tools and section~\ref{sec:fitting} demonstrates fitting with the full forward-model. Section~\ref{sec:implementation} gives details of the software implementation, and Section~\ref{sec:summary} summarises the NeSST model.

\section{\label{sec:primary}Primary Neutron Spectra}

The three thermonuclear reactions of primary interest in DT fusion experiments are:
\begin{align}
  D + T &\to \alpha + n,
    & \bar{E}_n \approx 14.03~\mathrm{MeV},
    \label{eq:DTreac}\\[2pt]
  D + D &\to {}^3\mathrm{He} + n,
    & \bar{E}_n \approx 2.45~\mathrm{MeV},
    \label{eq:DDreac}\\[2pt]
  T + T &\to \alpha + 2n,
    &  .
    \label{eq:TTreac}
\end{align}
The DT and DD reactions are two-body and produce near-Gaussian\cite{Ballabio1998} neutron peaks centred near 14~MeV and 2.45~MeV respectively, broadened by thermal and non-thermal Doppler motion to typical widths of order hundreds of keV.  The TT reaction is a three-body process producing a broad, continuous spectrum spanning the full kinematically allowed range from 0 to $\approx 10$~MeV. 

\subsection{DT and DD spectra}
\subsubsection{Spectral moments: Ballabio fits}

As both DT and DD are two-body reactions with two products, reaction kinematics can be used to determine the product energies precisely. Spectral-moment theories have been developed to describe these product spectra by exploiting the fact that fusion products are much more energetic than the reacting ions. A full review of spectral-moment theory is beyond the scope of this work; relevant treatments are given by Brysk\cite{Brysk1973}, Ballabio\cite{Ballabio1998}, Munro\cite{Munro2016,Munro2017}, Appelbe\cite{appelbe2024generalizing}, and Crilly\cite{crilly2022constraints}. In short, the shape of the DT and DD primary spectra can be summarised by a set of cumulants which can be related to burn-averaged plasma conditions.

For a static fusing plasma, the mean energy $\bar{E}$ and variance $\sigma_E^2$ of the distribution of neutrons emitted in a thermonuclear reaction depend on the ion temperature $T_i$.  NeSST uses the fits of Ballabio et~al.~\cite{Ballabio1998}, which capture the leading-order Doppler broadening and higher-order corrections to the centroid energy for DT and DD reactions.  The DT centroid shift from the zero-temperature Q-value $E_0^{DT}$ is
\begin{equation}
  \Delta\bar{E}_{DT} = a_1 T_i^{2/3}(1 + a_2 T_i^{a_3})^{-1} + a_4 T_i,
\end{equation}
where $T_i$ is in keV and the coefficients $\{a_1,a_2,a_3,a_4\}$ are tabulated in Table~III of Ref.~\cite{Ballabio1998} (applicable in the range 0 $< T_i <$ 40 keV).  Similarly, the variance is
\begin{equation}
  \sigma_E^2 = \frac{C^2 T_i}{(2\sqrt{2\ln 2})^2}, \qquad
  C = \omega_0(1+\delta_\omega(T_i)),
\end{equation}
where $\omega_0$ and $\delta_\omega(T_i)$ encode relativistic first-order and higher-order correction terms respectively.  Corresponding formulae hold also for DD. Simpler expressions for these moments can be found in Brysk\cite{Brysk1973}, but these are not relativistically correct.

\subsubsection{Spectral shapes}

The spectral moments are summary statistics of the full neutron spectrum. However, there is no simple mapping from these moments to the accurate spectral shape. Instead we use a set of approximate functional forms and validate these against Monte Carlo calculations.

The simplest approximation for the primary spectrum is a Gaussian centred on the shifted mean:
\begin{equation}
  Q_\mathrm{Brysk}(E) =
    \frac{1}{\sqrt{2\pi\sigma_E^2}}
    \exp\!\left(-\frac{(E-\bar{E})^2}{2\sigma_E^2}\right).
\label{eq:brysk}
\end{equation}
This is the Brysk approximation~\cite{Brysk1973}.

Ballabio et~al.~\cite{Ballabio1998} derived a modified Gaussian that better captures the asymmetry of the neutron spectrum arising from relativistic kinematics.  NeSST implements equations~(44)--(47) of that reference:
\begin{equation}
  Q_\mathrm{Ballabio}(E) = \frac{1}{\sqrt{2\pi\sigma_E^2}}
    \exp\!\left(
      -\frac{2\bar{E}_\mathrm{eff}(\sqrt{E}-\sqrt{\bar{E}_\mathrm{eff}})^2}
            {\sigma_\mathrm{eff}^2}
    \right),
\end{equation}
where $\bar{E}_\mathrm{eff}$ and $\sigma_\mathrm{eff}$ are scaled mean and width parameters defined in Ref.\cite{Ballabio1998}.

For reference calculations that make no approximation about the spectral shape, NeSST provides a wrapper to pyDRESS~\cite{eriksson2016dress}, a Monte Carlo code that samples the fusion kinematics directly.  This is particularly useful for quantifying the accuracy of the analytical shapes -- a comparison is shown in Fig. \ref{fig:primary}.

\subsection{TT spectrum}

The TT reaction $\mathrm{T+T\to{}^4He+2n}$ produces a continuous neutron spectrum that cannot be represented as a simple two-body peak. To accurately capture the shape, a R-matrix formalism was developed by Brune \textit{et al.}\cite{brune2015}. It describes the reaction in terms of partial waves, each corresponding to a different angular momentum state of the interacting nuclei and decomposes the cross-section into their respective contributions. These contributions are quantified by feeding factors, which can be inferred by fitting the neutron data obtained in experiment. Dedicated inertial\cite{brune2015,GatuJohnson2018} and magnetic\cite{eriksson2024} confinement fusion experiments have been used to infer the value of these R-matrix feeding factors. The fitted R-matrix model can then be used to produce the centre-of-mass (CoM) frame TT neutron energy spectrum. NeSST adopts the Doppler-broadening formalism of Appelbe~et~al.~\cite{appelbe2016TT} to convert the CoM spectrum to the lab frame. This involves the convolution with a thermal Doppler kernel at a given ion temperature $T_i$:
\begin{align}
  \frac{dN}{dE}(E; T_i) &\propto \int dE_\mathrm{CoM}\;
    \frac{\phi(E_\mathrm{CoM})}{\sqrt{E_\mathrm{CoM}}} \label{eq:TT_broadening} \ \\
    &\sqrt{\frac{m_T}{2 \pi m_n T_i}}
    \exp\!\left(-\frac{2m_T}{m_n T_i}
           \left(\sqrt{E}-\sqrt{E_\mathrm{CoM}}\right)^2\right), \nonumber
\end{align}
where $\phi(E_\mathrm{CoM})$ is the normalised CoM spectral shape and $m_T$ is the triton mass. In NeSST we assume the CoM spectrum is independent of reaction energy, although Gatu-Johnson \textit{et al.} shows experimental evidence that this assumption may be incorrect.

NeSST provides several choices of CoM spectral model, all stored as pre-normalised tabulated distributions and selected via the \texttt{model} argument of \texttt{dNdE\_TT}. These currently include Brune's fit 16\cite{brune2015}, three fits at different reaction energies from Gatu-Johnson\cite{GatuJohnson2018} and the magnetic confinement fusion Eriksson \textit{et al.} fit\cite{eriksson2024}. All models are convolved with the same Doppler-broadening kernel, Eq.~(\ref{eq:TT_broadening}). Figure~\ref{fig:primary}(d) compares the broadened TT spectra at $T_i = 5$~keV across all available models.

\subsection{Yields and reactivities}

\begin{figure}[t]
  \centering
  \includegraphics[width=\columnwidth]{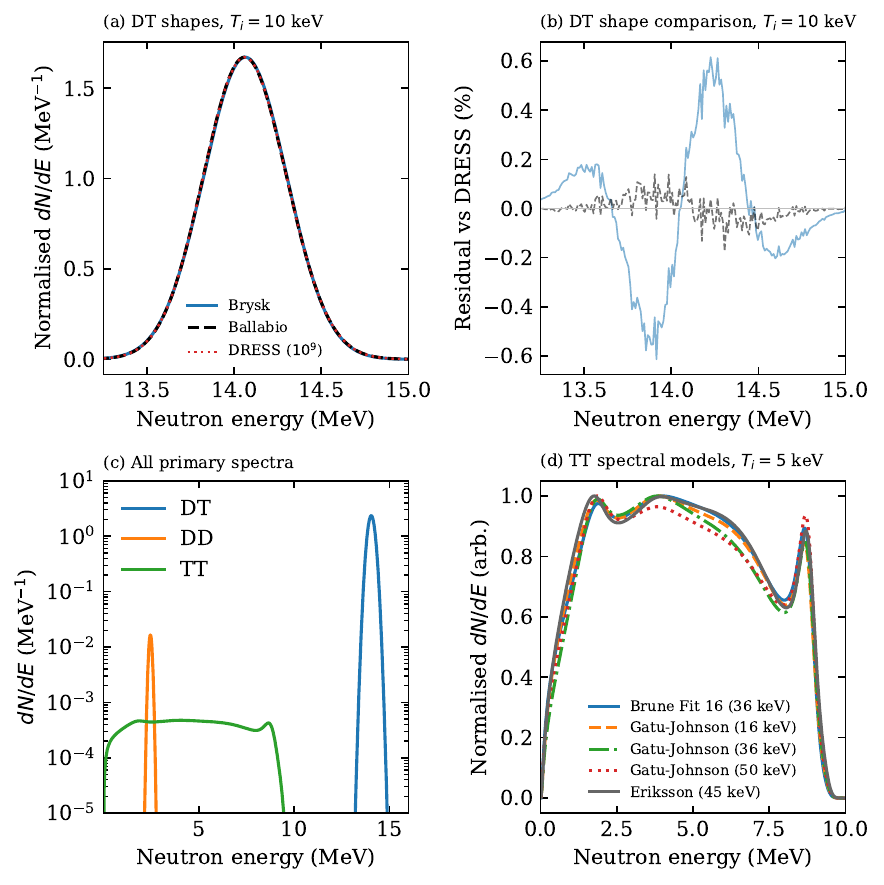}
  \caption{%
    (a) The DT spectrum using various spectral shape models at 10 keV.
    (b) Residual between approximate spectral shape models and DRESS Monte Carlo ($10^9$ samples) shapes for the DT primary spectrum.
    (c) All primary spectra using default models from equimolar 5 keV DT plasma.
    (d) TT spectral models at $T_i=5$~keV.  All spectra are normalised to their
    peak value. Numbers in brackets are reported reaction energies in keV.
  }
  \label{fig:primary}
\end{figure}

For plasmas in local thermal equilibrium, fusion yield ratios between reactions can be estimated from thermonuclear reactivities.  NeSST provides a number of different parametrisations for reactivities. The defaults are the Bosch--Hale parametrisations~\cite{bosch1992} for $\langle\sigma v\rangle_{DT}$ and $\langle\sigma v\rangle_{DD}$, and numerical integration of ENDF data for $\langle\sigma v\rangle_{TT}$.  Defining the volumetric reaction rate as $\dot{n}_{ij} = n_i n_j \langle\sigma v\rangle_{ij} / (1+\delta_{ij})$ (where the Kronecker factor $\delta_{ij}$ prevents
double-counting identical reactants), the DD and TT neutron yields normalised to the DT yield are:
\begin{equation}
  \frac{Y_{DD}}{Y_{DT}} =
    \frac{\tfrac{1}{2}f_D^2\,\langle\sigma v\rangle_{DD}}
         {f_D f_T\,\langle\sigma v\rangle_{DT}},
  \quad
  \frac{Y_{TT}}{Y_{DT}} =
    \frac{f_T^2\,\langle\sigma v\rangle_{TT}}
         {f_D f_T\,\langle\sigma v\rangle_{DT}},
\end{equation}
where $f_D$ and $f_T$ are the deuterium and tritium number fractions, the factor of $1/2$ for DD accounts for identical reactants, and the numerator for TT already accounts for the factor of two neutrons produced per TT reaction.  Figure~\ref{fig:primary} shows representative primary spectra for all three reactions at $T_i=5$~keV, together with a comparison of the DT spectral shapes.

For the majority of data analysis and synthetic data generation, using a single (apparent\cite{GatuJohnson2018}) ion temperature to determine spectral shapes and yield ratios is sufficient. However, in reality the fusion plasma has a distribution of temperatures and fluid velocities which gives rise to corrections to the spectral shape\cite{Munro2016} and yield ratios\cite{kabadi2021} over simple averages. To include these corrections, one can define a joint probability distribution function (PDF) over ion temperature and fluid velocity and perform summations over it. This PDF is equivalent to a neutron- or burn-weighting. For example, the total DT neutron spectrum can be computed as follows:
\begin{subequations}
\begin{align}
    \left.\frac{dN}{dE}\right|_{\mathrm{DT, total}} &= Y_{DT} \int  \int  P(T_i, v_{\parallel}) \frac{dN}{dE}(T_i, v_{\parallel}) \ dT_i dv_{\parallel} \ , \\
    \frac{dN}{dE}(T_i, v_{\parallel}) &= Q_{\mathrm{Ballabio}}\left(\bar{E}_{DT}(T_i, v_{\parallel}), \sigma_{DT} (T_i) \right) \ , 
\end{align}
\end{subequations}
where $v_{\parallel}$ is the fluid velocity projected to the detector line of sight and is included as a Doppler shift to the mean energy. Note that using a burn-weighting does not account for neutron transport effects, see Munro\cite{Munro2016} for a detailed discussion of this. Also the spectral shape does not sufficiently constrain the PDF, thus the inverse problem is highly degenerate. As this analysis is more specialised, it is not included as a core functionality of NeSST, but can be implemented by using the above formulae.

\section{\label{sec:scatter}Single-Scatter Spectral Model}

Scattered neutron spectra in ICF depend on the primary reaction spectra, nuclear differential cross sections, and the geometry of the scattering medium. As the primary fusion neutrons are emitted isotropically this allows simplification in the calculation, effectively reducing the evaluation to reductions over the scattering matrices. The following approximations allow NeSST to accurately compute scattered spectra without the need for solving the full neutron transport equation.

\subsection{Scattering geometry and areal density}

The scattered neutron spectrum for a single scattering species $i$ is obtained by convolving the primary spectrum with the double differential nuclear cross section, weighted by the neutron-averaged projected areal density along each scattering chord.  Following Ref.~\cite{Crilly2021}, we define the neutron-averaged areal density seen along a scattering direction at cosine $\mu_s$ and azimuth $\phi_s$ as
\begin{equation}
  \langle\rho L\rangle(\mu_s, \phi_s) =
    \frac{1}{Y_n}\int_V d^3\mathbf{r}
    \int_0^\infty \rho(\mathbf{r}+s\hat{\Omega})\,ds\; R_n(\mathbf{r}),
\label{eq:rhoL}
\end{equation}
where $\hat{\Omega}$ is the scattering direction at angle $\theta_s = \cos^{-1}(\mu_s)$ to the detector, $R_n$ is the neutron production rate and $Y_n$ is the total neutron yield.  Due to the azimuthal symmetry of the differential cross sections for the nuclear processes of interest, it is natural to integrate out the azimuth and expand $\langle\rho L\rangle(\mu_s)$ in Legendre polynomials:
\begin{equation}
  \langle\rho L\rangle(\mu_s) =
    \sum_{l=0}^{\infty} \langle\rho L\rangle_l \, P_l(\mu_s).
\label{eq:rhoL_legendre}
\end{equation}
The $l=0$ coefficient $\langle\rho L\rangle_0$ is the $4\pi$-averaged areal density and sets the overall scattering amplitude.  Higher-order coefficients encode asymmetries. It should be noted that a single detector line of sight integrates over azimuthal angle and is therefore insensitive to spherical-harmonic modes $Y_l^m$ with $m\ne0$; reconstructing full non-uniformities requires multiple lines of sight~\cite{Crilly2021}.

Finally, each scattering component is weighted by an amplitude based on the $4\pi$-averaged areal density and the cross section:
\begin{equation}
  A_{1S} = \frac{\langle\rho L\rangle_0}{\bar{m}}\,\bar{\sigma},
\label{eq:A1s}
\end{equation}
where $\bar{m}$ is the mean ion mass and $\bar{\sigma}$ is a defined cross section scaling factor (taken as 1 barn).  The quantity $A_{1S}$ provides a convenient dimensionless scaling such that all spectral shapes can be pre-computed and subsequently multiplied by $A_{1S}$ when evaluating at a specific areal density.

\subsection{Symmetric (isotropic) areal density}

For a spherically symmetric implosion ($\langle\rho L\rangle = \langle\rho L\rangle_0 = \text{const}$), the scattered spectrum from species $i$ is
\begin{equation}
  \left.\frac{dN}{dE}\right|_\mathrm{scat}^{(i)} =
    A_{1S}\int dE'\,\frac{d\sigma_i}{dE}(E',E)\,
    \frac{dN}{dE'}\bigg|_\mathrm{prim},
\label{eq:sym_scatter}
\end{equation}
where $d\sigma_i/dE$ is the single differential cross section obtained by integrating the double differential cross section over all scattering angles.

\subsection{Asymmetric areal density}

In the presence of areal density asymmetries, the full double differential cross section must be retained and the asymmetry function integrated over scattering angle:
\begin{align}
  \left.\frac{dN}{dE}\right|_\mathrm{scat}^{(i)} &=
    A_{1S}\int_{-1}^{1} d\mu_s\,
    \frac{\langle\rho L\rangle(\mu_s)}{\langle\rho L\rangle_0} \nonumber \\
    &\int dE'\,\frac{d^2\sigma_i}{dEd\mu_s}(E',E,\mu_s)\,
    \frac{dN}{dE'}\bigg|_\mathrm{prim}.
\label{eq:asym_scatter}
\end{align}
The ratio $\langle\rho L\rangle(\mu_s)/\langle\rho L\rangle_0$ can be expressed as the Legendre expansion, Eq.~(\ref{eq:rhoL_legendre}), and/or supplied by the user as a general callable function of $\mu_s$, normalised such that the zeroth Legendre coefficient is unity.

\begin{figure}[h]
  \centering
  \includegraphics[width=0.8\columnwidth]{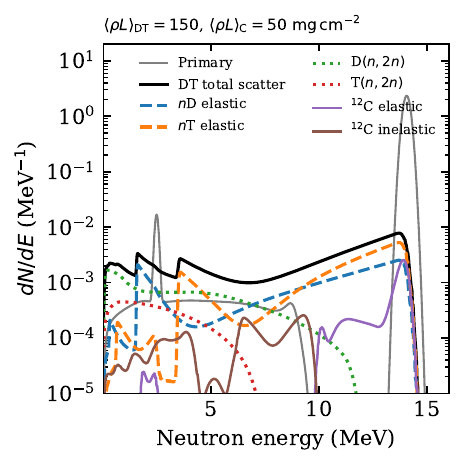}
  \caption{%
    Scattered spectrum components for an equimolar DT plasma at
    $T_i=5$~keV, with isotropic scattering from $\langle\rho L\rangle_\mathrm{DT} =
    150~\text{mg/cm}^2$, together with the $^{12}$C ablator
    contribution at $\langle\rho L\rangle_\mathrm{C} =
    50~\text{mg/cm}^2$.  Kinematic backscatter edges of the elastic
    $n$D and $n$T components lie near 1.56 and 3.5~MeV for
    stationary ions (see Sec.~\ref{sec:ionkin}).
  }
  \label{fig:components}
\end{figure}

\subsection{Nuclear cross sections}

\subsubsection{Elastic scattering}

Total and differential elastic cross sections for $n$D, $n$T and other species are read from Evaluated Nuclear Data File (ENDF)\cite{ENDF} libraries using an in-built ENDF interface (making use of the endf-python package).  The differential cross section in the centre-of-mass (CoM) frame is represented either as a Legendre polynomial expansion,
\begin{equation}
  \frac{d\sigma}{d\Omega_\mathrm{CoM}}(E_i, \mu_c)
  = \frac{\sigma_\mathrm{el}(E_i)}{4\pi}
    \sum_{l=0}^{N_l} \frac{2l+1}{2} a_l(E_i) P_l(\mu_c),
\end{equation}
where the Legendre coefficients $a_l(E_i)$ are interpolated from tabulated ENDF data, or as a tabulated angular distribution for data provided in that format.

The CoM scattering cosine $\mu_c$ is related to the lab-frame incoming and outgoing energies $(E', E)$ through the scattering kinematics. NeSST implements both classical and relativistic kinematics, with relativistic corrections enabled by default via the Lorentz-invariant four-momentum formalism.  The Jacobian $\partial\mu_c/\partial E$ is computed analytically, allowing the conversion of CoM frame single differential cross sections to lab-frame $d\sigma/dE$ as a matrix in $(E',E)$ space.

\subsubsection{Inelastic ($n,2n$) processes}

The reactions D$(n,2n)$p and T$(n,2n)$D are three-body breakup processes that produce a broad, continuous spectrum of outgoing neutrons within DT fuel.  NeSST supports two data formats for the double differential cross section of these processes:
\begin{itemize}
  \item \textbf{LAW~6 (phase space law)}: used for T$(n,2n)$D; the
    isotropic phase-space distribution of Ref.~\cite{ENDF} is
    used to construct the outgoing energy-angle distribution.  This simple model represents
    a large uncertainty in spectral shape.
  \item \textbf{LAW~7 (tabulated)}: used for D$(n,2n)$p; the
    double differential cross section tabulated in CENDL~\cite{cendl}
    is used directly.  This cross section has been compared against
    the Deltuva calculation\cite{deltuva2003} and Forrest measurement\cite{Forrest2019} and shows good agreement.
\end{itemize}

These double differential cross sections are used to construct scattering matrices of the same energy dimension as the single differential cross sections used for the elastic processes.

\subsection{Total scattered spectrum and components}

The total scattered spectrum from a DT implosion is the sum of contributions from all scattering processes weighted by the species fractions $f_D$ and $f_T$:
\begin{align}
  \frac{dN}{dE}\bigg|_\mathrm{scat}
  &= f_D\left(\frac{dN}{dE}\bigg|_{nD}
            + \frac{dN}{dE}\bigg|_{D(n,2n)}\right) \nonumber \\
  &+ f_T\left(\frac{dN}{dE}\bigg|_{nT}
            + \frac{dN}{dE}\bigg|_{T(n,2n)}\right).
\label{eq:total_scatter}
\end{align}
Additional species, such as $^{12}$C from a high density carbon ablator, are included through separate calls to \texttt{mat\_scatter\_spec}. Figure~\ref{fig:components} shows the individual DT components alongside the $^{12}$C elastic and inelastic contributions for a representative case.

\subsection{Effect of areal density asymmetry}

\begin{figure}[t]
  \centering
  \includegraphics[width=0.8\columnwidth]{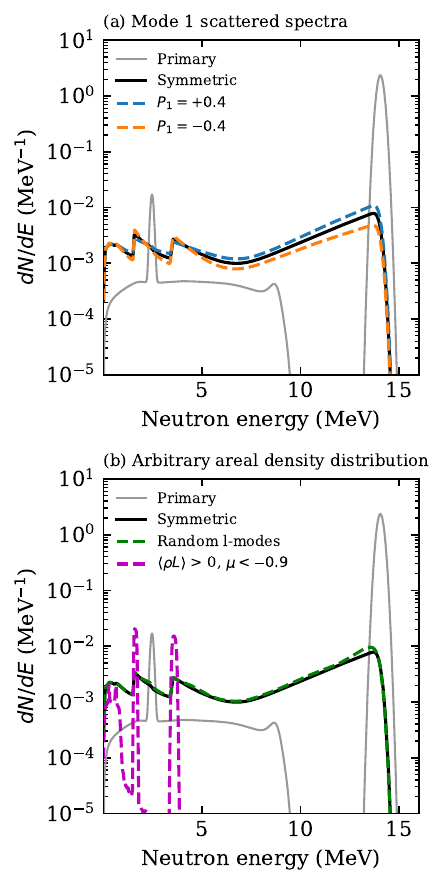}
  \caption{%
    (a) Scattered spectra for symmetric ($P_1=0$, black) and
    mode-1 asymmetric ($P_1=\pm 0.4$, coloured) areal density
    distributions at $\langle\rho L\rangle_0=150~\text{mg/cm}^2$.
    (b) Arbitrary areal density distributions including random Legendre modes and a block of scattering material at angles greater than $>$ 155$^o$, both at $\langle\rho L\rangle_0=150~\text{mg/cm}^2$.
  }
  \label{fig:asymmetry}
\end{figure}

By using Eq.~\ref{eq:asym_scatter}, one can compute the singly scattered neutron spectrum from asymmetric distributions of scattering media. To do this, one must additionally provide the areal density as a function of scattering cosine. 

Figure~\ref{fig:asymmetry}(a) illustrates the effect of a mode-1 areal density asymmetry, $\langle\rho L\rangle(\mu_s) = \langle\rho L\rangle_0(1+P_1\mu_s)$, on the scattered spectrum.  The leading sensitivity to $P_1$ arises through the angular weighting of the double differential cross section: forward-scattered neutrons ($\mu_s \to 1$) lose little energy and have large cross section, while backward-scattered neutrons ($\mu_s \to -1$) lose maximum energy. An excess (deficit) in the forward-facing areal density therefore raises (lowers) the high-energy shoulder of the elastic scattering continuum relative to the symmetric case. Additionally, in figure~\ref{fig:asymmetry}(b) we show the flexibility to more arbitrary areal density distributions including random Legendre modes and a block of scattering material at angles greater than 155$^\circ$.

\section{\label{sec:ionkin}Ion Kinematics and the Backscatter Edge}

\subsection{Physical picture}

The kinematic lower bound of 180$^\circ$ elastic neutron scattering produces a sharp ``backscatter edge'' in the neutron spectrum.  For $n$T ($n$D) scattering of a DT primary neutron, the edge energy for a stationary triton (deuteron) is given in the non-relativistic limit by
\begin{equation}
  E_\mathrm{edge} = \left(\frac{A-1}{A+1}\right)^2 E_\mathrm{prim},
\label{eq:edge}
\end{equation}
where $A = m_i/m_n$ is the ion-to-neutron mass ratio.  Using the physical mass ratios $A_T = m_T/m_n \approx 3$ and $A_D = m_D/m_n \approx 2$, and $E_\mathrm{prim}\approx 14.03$~MeV for a stationary DT plasma, this gives $E_{nT}\approx 3.5$~MeV and $E_{nD}\approx 1.56$~MeV. NeSST applies relativistic kinematics by default, which shifts these edges by a small but non-negligible amount; the non-relativistic formula is quoted here for orientation. 

If the scattering ions have a velocity distribution, each ion velocity $v_{i,\parallel}$ (parallel to the neutron direction) shifts the edge energy for that interaction.  The resulting backscatter edge shape is therefore a convolution of the ion velocity distribution with the single-ion edge shape~\cite{Crilly2020}:
\begin{equation}
  \frac{dN}{dE}\bigg|_\mathrm{bs} \propto \int dv_{i,\parallel}\,
    P(v_{i,\parallel})\,Q_{bs}(v_{i,\parallel}, E),
\label{eq:Ibs}
\end{equation}
where $P(v_{i,\parallel})$ is the scattering-rate-weighted ion velocity distribution and $Q_{bs}$ is the single-velocity edge shape determined by the primary spectrum, differential cross section and Doppler shift due to $v_{i,\parallel}$.

A bulk flow velocity $\bar{v}$ shifts the edge position linearly, while the variance of the velocity distribution (from ion temperature or fluid velocity variance) broadens it.  Because the velocity distribution $P(v_{i,\parallel})$ is weighted by the local neutron scattering rate, the inferred moments reflect the areal-density-weighted temperature and flow, not the hotspot.  This makes the backscatter edge a diagnostic of hydrodynamic conditions in the dense shell~\cite{Crilly2020,Crilly_POP2024}.

\begin{figure}[t]
  \centering
  \includegraphics[width=0.8\columnwidth]{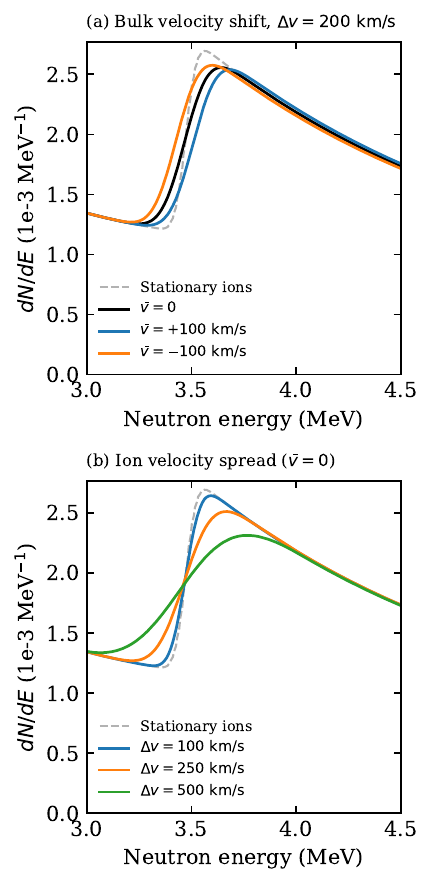}
  \caption{%
    The $n$T backscatter edge at
    $\langle\rho L\rangle=150~\text{mg/cm}^2$, $T_i=5$~keV.
    (a) Effect of bulk ion velocity: a positive (negative) bulk
    velocity shifts the edge to higher (lower) energies.  The dashed
    line shows the stationary-ion result.
    (b) Effect of ion velocity spread (temperature) at zero bulk
    velocity: increasing $\Delta v$ broadens the edge.
  }
  \label{fig:edge}
\end{figure}

\subsection{Implementation}

NeSST pre-computes the ion-velocity-dependent scattering kernel on a three-dimensional grid in $(E_\mathrm{out}, v_i, E_\mathrm{in})$ space, using the elastic scattering energy-angle relation to find the scattering cosine.  During spectrum evaluation, the kernel is integrated against a specified ion velocity distribution—by default a Gaussian characterised by mean $\bar{v}$ and standard deviation $\Delta v$—via reduction over the full scattering matrix, allowing rapid evaluation at many $(\bar{v}, \Delta v)$ values during fitting.  The velocity-independent (stationary-ion) spectrum can be recovered as a special case, but is instead computed using the scattering matrices described in Section~\ref{sec:scatter}.

Figure~\ref{fig:edge} shows the effect of bulk velocity and velocity spread on the nT backscatter edge shape.

\section{Higher areal density effects}\label{sec:highrhoL}

While NeSST is built around the single scatter approximation and the approximate isotropy of the primary spectrum, there are methods to extend NeSST to higher areal densities. To do this, one needs to capture multi-scatter effects, to leading order these are double scattering and transmission.

\subsection{Double scatter and fuel transmission}

At higher areal densities ($\langle\rho L\rangle \gtrsim 100~\text{mg/cm}^2$) a non-negligible fraction of neutrons undergo two (or more) collisions before leaving the fuel and reaching the detector. Higher-order scattering is complicated by the fact that the flux of singly scattered neutrons is anisotropic, with strong energy-angle correlations due to elastic scattering. Proper accounting of these anisotropy effects and the influence on fuel geometry requires dedicated transport codes.  As an estimate of the second-scatter contribution, NeSST offers an approximate formula obtained by re-applying the single-scatter kernel to the first-scatter spectrum:
\begin{equation}
  \left.\frac{dN}{dE}\right|_{\mathrm{2nd}} \approx
  A_{1S}^2
  \int dE'\,\frac{d\sigma}{dE}(E',E)\,
  \left.\frac{dN}{dE'}\right|_{\mathrm{1st}},
\label{eq:double_scatter}
\end{equation}

Additionally, the fuel areal density also attenuates the primary neutron spectrum via a Beer--Lambert law:
\begin{equation}
  \mathcal{T}(E) =
    \exp\!\left[
      -A_{1S}\,\hat{\rho L}(\hat{\Omega}_\mathrm{det})
      \bigl(f_D\,\sigma_D^\mathrm{tot}(E)
           +f_T\,\sigma_T^\mathrm{tot}(E)\bigr)
    \right],
\label{eq:transmission}
\end{equation}
where $\hat{\rho L}(\hat{\Omega}_\mathrm{det})$ is the normalised areal density along the detector line of sight. Similarly, the single scatter signal is attenuated when multi-scattering is non-negligible, although here the areal density needed to compute the transmission factor is difficult to obtain due to scattering geometry. It is therefore clear that while some extensions to NeSST's single scatter basis are possible, they require additional assumptions which are more difficult to justify. In the following section, we use comparison to 1D neutron transport to provide an a posteriori assessment of the approximate double scatter model.

\subsection{Comparison with 1-D neutron transport}

\begin{figure}[htp]
  \centering
  \includegraphics[width=1.05\columnwidth]{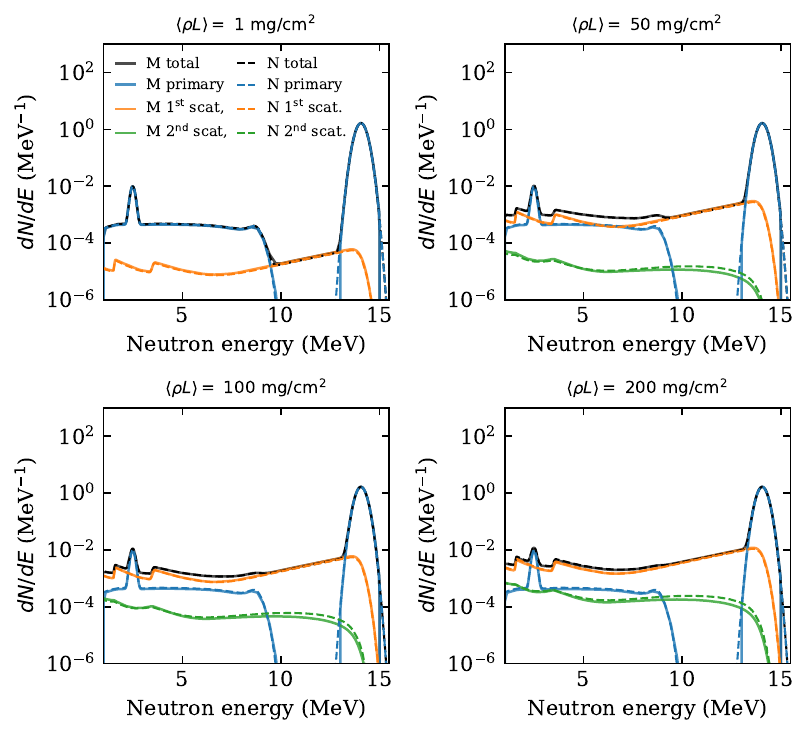}
  \caption{%
    Comparison of NeSST (dashed, N) with Minotaur 1-D transport (solid, M)
    spectra for a simple ice block model at various
    $\langle\rho L\rangle$.  NeSST reproduces the primary and single-scatter components well when compared to the full transport result. The approximate double scatter model also shows qualitative agreement, although shows larger error than the other components.
  }
  \label{fig:minotaur}
\end{figure}

Figure~\ref{fig:minotaur} compares NeSST spectra with those from Minotaur~\cite{Crilly2018}, a 1-D spherical, multi-group, discrete ordinates neutron transport code written in Fortran 90, for a simple ice block model at various $\langle\rho L\rangle$ between 1  and 200 mg/cm$^2$. 

The setup time of the scattering matrices between the codes was similar ($\sim$ 15s), however the time to compute a neutron spectrum after setup was substantially different, NeSST taking 2.5 s and Minotaur taking 60 s. This comparison was for 600 neutron energy groups in NeSST and 100 in Minotaur.

The primary spectra and single-scatter components agree well across the full energy range.  Systematic differences arise when comparing the double scatter spectra, showing that NeSST overestimates the degree of double scatter at higher neutron energies. This is to be expected as the high energy doubly scattered neutrons have undergone two small angle scattering events, thus experiencing lower average areal density when compared to the large scattering angle trajectories.

Listing~\ref{lst:DT} shows how to compute the full energy spectrum, including DT, DD, TT primaries, and single and double scattering, for a symmetric areal density configuration. 
\begin{lstlisting}[caption={Computing the full energy spectrum for a symmetric DT plasma.},label={lst:DT}]
import NeSST as nst
import numpy as np

E_arr = np.linspace(1e6, 16e6, 500)   # eV
rhoL = 1.5 # kg/m^2
T_ion = 10.0e3 # eV
frac_D = 0.5
frac_T = 0.5

Y_DT, Y_DD, Y_TT = nst.yields_normalised(T_ion, frac_D=frac_D, frac_T=frac_T)

DTmean, _, DTvar = nst.DTprimspecmoments(T_ion)
DDmean, _, DDvar = nst.DDprimspecmoments(T_ion)

A_1S = nst.rhoR_2_A1s(rhoL)

nst.init_DT_scatter(E_arr, E_arr)
dNdE_prim = (
    Y_DT * nst.QBallabio(E_arr, DTmean, DTvar)
    + Y_DD * nst.QBallabio(E_arr, DDmean, DDvar)
    + Y_TT * nst.dNdE_TT(E_arr, T_ion)
)
dNdE_scatter, _ = nst.DT_sym_scatter_spec(dNdE_prim)
dNdE_double_scatter, _ = nst.DT_sym_scatter_spec(dNdE_scatter)

dNdE_total = dNdE_prim + A_1S * dNdE_scatter + A_1S**2 * dNdE_double_scatter
\end{lstlisting}

\section{\label{sec:materials}Additional Scattering Materials}

Beyond the DT fuel, ablator and structural materials can scatter fusion neutrons and contribute to the measured spectrum.  NeSST provides a general \texttt{material\_data} class that loads ENDF-format cross section data from JSON configuration files bundled with the package.  Materials currently bundled with NeSST include H, D, T, $^{12}$C, $^9$Be, and others accessible from the ENDF library directory (see the package documentation for the current list).  Each material supports elastic scattering with Legendre-coefficient or tabulated angular distributions; discrete-level inelastic scattering; and $(n,2n)$ processes where present in the data. Note that not all scattering laws implemented in ENDF have been implemented in NeSST, the focus is those relevant to DT fusion plasmas.  The \texttt{mat\_scatter\_spec} function computes the singly scattered spectrum from any bundled material in a form equivalent to Eq.~(\ref{eq:asym_scatter}).  Support for additional materials requires providing appropriate JSON configuration and ENDF files, which can then be loaded through the ENDF interface.

Listing~\ref{lst:materials} shows how to compute the scattered spectrum from $^{12}$C. Figure~\ref{fig:components} shows a calculated singly scattered spectrum from $^{12}$C.

\begin{lstlisting}[caption={Computing the scattered spectrum from $^{12}$C ablator material.},label={lst:materials}]
import NeSST as nst
import numpy as np

E_arr = np.linspace(1e6, 16e6, 500)   # eV
T_ion = 5e3 # eV
rhoL_C12 = 5.0 # kg/m^2

def symmetric_rhoL(mus):
    return np.ones_like(mus)

# Primary spectrum (e.g. DT Brysk)
mean, _, var = nst.DTprimspecmoments(T_ion)
I_E = nst.QBrysk(E_arr, mean, var)

# Initialise C12 and compute scattered spectrum
mat_C12 = nst.init_mat_scatter(E_arr, E_arr, 'C12')
dNdE_C12 = nst.mat_scatter_spec(mat_C12, I_E, symmetric_rhoL)
A_1S = mat_C12.rhoR_2_A1s(rhoL_C12)
scattered_spectrum = A_1S * dNdE_C12
\end{lstlisting}

\section{\label{sec:ntof}Synthetic Neutron Time-of-Flight Diagnostics}

Neutron time-of-flight (nToF) spectrometers are one of the primary experimental methods for neutron spectroscopy at ICF facilities. Following Hatarik~et~al.~\cite{hatarik2015} and Mohamed~et~al.~\cite{mohamed2020}, the canonical forward model of nToF spectrometers posits that the measured signal is related to the neutron energy spectrum via:
\begin{align}
  I(t_d) &\propto \int dt_a\;
    \underbrace{S(E_n)}_{\text{sensitivity}}
    \cdot\underbrace{\mathcal{T}(E_n)}_{\text{LOS atten.}}
    \cdot\underbrace{\frac{dN}{dE_n}}_{\text{spectrum}} \nonumber \\
    &\cdot\underbrace{\left|\frac{dE_n}{dt_a}\right|}_{\text{Jacobian}}
    \cdot\underbrace{R(E_n,\, t_d - t_a(E_n))}_{\text{IRF}},
\label{eq:ntof}
\end{align}
where $t_a(E_n) = d/v_n(E_n)$ is the neutron arrival time at distance $d$, $S(E_n)$ is the energy-dependent scintillator sensitivity, $\mathcal{T}(E_n)$ is the line-of-sight transmission/attenuation, and $R(E_n, t_d-t_a)$ is the energy-dependent~\cite{mohamed2020} instrument response function (IRF).  The (relativistic) Jacobian, $|dE_n/dt_a| = m_n \beta^2 \gamma^3 / t_a$, converts between energy and arrival-time representations.  

NeSST makes use of the normalised time domain to simplify calculations, this converts the real detection time to one normalised by the light transit time to the detector:
\begin{equation}
    \tau = t \cdot \frac{c}{d} = \frac{1}{\beta} \ ,
\end{equation}
where $d$ is the detector distance and $\beta = v/c$ is the normalised velocity. From this we can see that all neutron signals must reach the detector at $\tau > 1$ and every detector distance has an equivalent signal in $\tau$ (up to IRF and finite burn duration effects).

\subsection{Instrument response functions}

NeSST constructs the IRF as an $(N\times N)$ matrix $\mathbf{R}$
whose entry $R_{jk}$ gives the fractional signal detected at time
$t_j$ from a mono-energetic neutron source arriving at time $t_k$.  A
composite IRF is built from a \emph{base} (or `neutron') response matrix (by default the neutron transit-time broadening through the scintillator volume) convolved column-wise with
a kernel function (known as the `X-ray' IRF). This separation of neutron and X-ray IRFs is well established\cite{hatarik2015}. NeSST provides the following building blocks for both neutron and X-ray IRFs:

\paragraph{Neutron IRF}
For a scintillator of thickness $\ell$, a neutron of speed $v_n$
spends $\Delta t = \ell/v_n$ inside the detector.  The canonical
top-hat base matrix assigns equal weight to all detection times within
this window:
\begin{equation}
  R^\mathrm{base}_{jk} \propto
    \mathbf{1}[t_k \le t_j \le t_k+\Delta t_k].
\end{equation}
This assumes the cross section for neutron interaction within the scintillator is very low and thus there is negligible scattering/attenuation of the neutron beam. However, for thicker scintillator volumes we expect non-negligible scattering and thus a reduction of flux with distance, as well as a long tail of interaction coming from multiple scattering of neutrons. 

The inverse-Gaussian neutron IRF model attempts to capture this behaviour and replaces the top-hat with a physically motivated model for scintillation in a thick organic scintillator. A neutron incident on a scintillator slab of thickness $\ell$ may interact on any pass through the material; multiple back-and-forth transits before escape give a distribution of interaction times that is approximately inverse-Gaussian in character.  The resulting base response has a prompt Beer--Lambert attenuation section followed by a heavy tail:
\begin{align}
 R^\mathrm{IG}(t; E_n) &\propto \Theta\left(t - \frac{\ell}{v_n}\right) e^{-f(E_n)\,v_n\,\sigma_p(E_n)\,n_p\,t} \label{eq:IG_nIRF}\\
    &+ A_{\mathrm{tail}}(E_n) \Theta\left(\frac{\ell}{v_n} - t\right) T_{IG}(\mu(E_n), \lambda(E_n); t) \nonumber \\
    T_{IG}(\mu, \lambda; t) &= \exp\left[\frac{\lambda}{\mu}\left(1-\sqrt{1+\frac{2 \mu^2 (t - \frac{\ell}{v_n})}{\lambda}} \right) \right] \nonumber
\end{align}
where $\Theta(x)$ is the Heaviside step function, $\sigma_p$ is the neutron--proton cross section, $n_p$ is the proton number density, and the Laplace-transformed inverse-Gaussian $T_{IG}(\mu,\lambda;t)$ characterises the signal associated with the first-hitting-time distribution of a one-dimensional random walk. Carbon or other scintillator constituents may also be included in the attenuation term. The energy-dependent coefficients $A_\mathrm{tail}$, $\mu$, $\lambda$ are fit to Monte Carlo calculations in 1 spatial dimension for a 10-cm organic scintillator -- one can fit this functional form to detailed calculations at a few different neutron energies and this model then provides a smooth interpolation.

\paragraph{X-ray IRF}
The temporal response of a neutron detector also includes light transport in the scintillator and any electronic response in the coupled detector. This response is typically measured directly from experiment using an intense, short flash of X-rays, hence the name ``X-ray'' IRF. The X-ray transit time is assumed to be negligible such that no correction needs to be made for this effect. The experimental measurements can then be fit to various functional forms. Murphy \textit{et al.}\cite{murphy1997} provides a number of simple analytic models for the temporal response for nToF detectors, these have been implemented in NeSST . A gated version of these IRF models is provided in NeSST:
\begin{align}
  \kappa(t) &\propto
    \frac{1+\mathrm{erf}\!\left(
      \frac{t-t_0-\sigma_g^2/\tau}{\sqrt{2}\,\sigma_g}\right)}{2\tau} \cdot \\
    &e^{-(t-t_0)/\tau+\sigma_g^2/(2\tau^2)}
    \cdot G\!\left(\frac{t-t_\mathrm{gate}}{\sigma_\mathrm{gate}}\right) \label{eq:kernel} \ , \\
    G(x) &= (1-e^{-x})/(1+e^{-x})\cdot\Theta(x) \ ,
\end{align}
where $\sigma_g$ is the Gaussian width of response, $\tau$ is the decay time, $t_0$ is a shift in the response centroid, and $G(x)$ is a one-sided logistic gate turn-on. The gate function was found to improve agreement between the model and experimental measurements (using $t_{gate} = 0$). The parameters $\{\sigma_g, \tau, t_0, \sigma_\mathrm{gate}\}$ are determined by fitting to X-ray pulse measurements on each shot campaign.

The composite IRF---inverse Gaussian neutron IRF $R^\mathrm{IG}(t; E_n)$ convolved with the gated decaying Gaussian X-ray IRF $\kappa(t)$---is illustrated in Fig.~\ref{fig:ntof}(c).

\subsection{Scintillator sensitivity}

The energy-dependent sensitivity of an organic scintillator at neutron energies below $\sim$10~MeV is dominated by the proton recoil process\cite{knoll2010}: an incident neutron transfers kinetic energy to hydrogen nuclei via $n$-$p$ scattering, and the recoiling protons generate scintillation light.  Tang~et~al.~\cite{tang2024} demonstrated a simple model of scintillator sensitivity, also known as non-linear light output (NLO), for a bibenzyl-based scintillator. Tang~et~al.~ showed that for isotropic recoil, the sensitivity is proportional to the light output integrated over all proton recoil energies:
\begin{equation}
  S(E_n) \propto \frac{\sigma_H(E_n)}{E_n}
    \int_0^{E_n} L(E_p)\,dE_p,
\label{eq:sensitivity}
\end{equation}
where $\sigma_H(E_n)$ is the $n$-H cross section and $L(E_p)$ is the proton light output. 

A number of models exist for $L(E_p)$ in the literature, including tabulated and simple semi-analytic functions. The simplest available model is a simple power law for $L(E_p) = E^p$\cite{knoll2010}. For tabulated data, Verbinski \textit{et al.}~\cite{verbinski1968} provides tabulated proton light output for NE-213 and NE-211 type plastic scintillators, this is interpolated and used in Eq~\ref{eq:sensitivity} within NeSST. Separately, Birks' law relates the light output to the stopping power ($dE/dx$):
\begin{equation}
    \frac{dL}{dE} \propto \frac{1}{1 + kB \frac{dE}{dx}} \ ,
\end{equation}
where $kB$ is a parameter which is material dependent. Similarly, Craun \& Smith\cite{craun1970} adds an extra term to this relation and shows improved fits to data:
\begin{equation}
    \frac{dL}{dE} \propto \frac{1}{1 + kB \frac{dE}{dx} + C \left(\frac{dE}{dx}\right)^2} \ ,
\end{equation}
Finally, these models can be closed using the Bethe (or Bethe-Bloch) stopping power model:
\begin{equation}
    \frac{dE}{dx} = \frac{a}{E} \log \left(\frac{4 m_e E}{m_p I}\right) \ ,
\end{equation}
where $a$ and the excitation energy, $I$, are parameters which are material specific. NeSST implements the above models, allowing flexibility in scintillator sensitivity modelling.

The appropriate model and its parameters are instrument-specific; neither the Verbinski nor the Birks parameterisation should be applied to a given detector without independent calibration. By default, NeSST normalises the scintillator sensitivity to be unity at the DT neutron energy.

\subsection{Line-of-sight attenuation}

Between target and detector, neutrons are attenuated by materials along the line of sight (air, windows, structural components). NeSST calculates the optical depth:
\begin{equation}
    \tau(E_n) = \sum_k n_k L_k \sigma_k^\mathrm{tot}(E_n) \ ,
\end{equation}
using ENDF total cross sections, giving a multiplicative transmission $\mathcal{T}(E_n) = e^{-\tau(E_n)}$. Comparisons against pencil-beam MCNP calculations for representative beamline configurations show good agreement with the transmission equation.

Figure~\ref{fig:ntof}(a) shows the computed transmission for 15~m of dry air at STP composed of $^{14}$N, $^{16}$O and $^{40}$Ar, as evaluated by NeSST.

\subsection{Full forward model and synthetic signal}

\begin{figure}[t]
  \centering
  \includegraphics[width=\columnwidth]{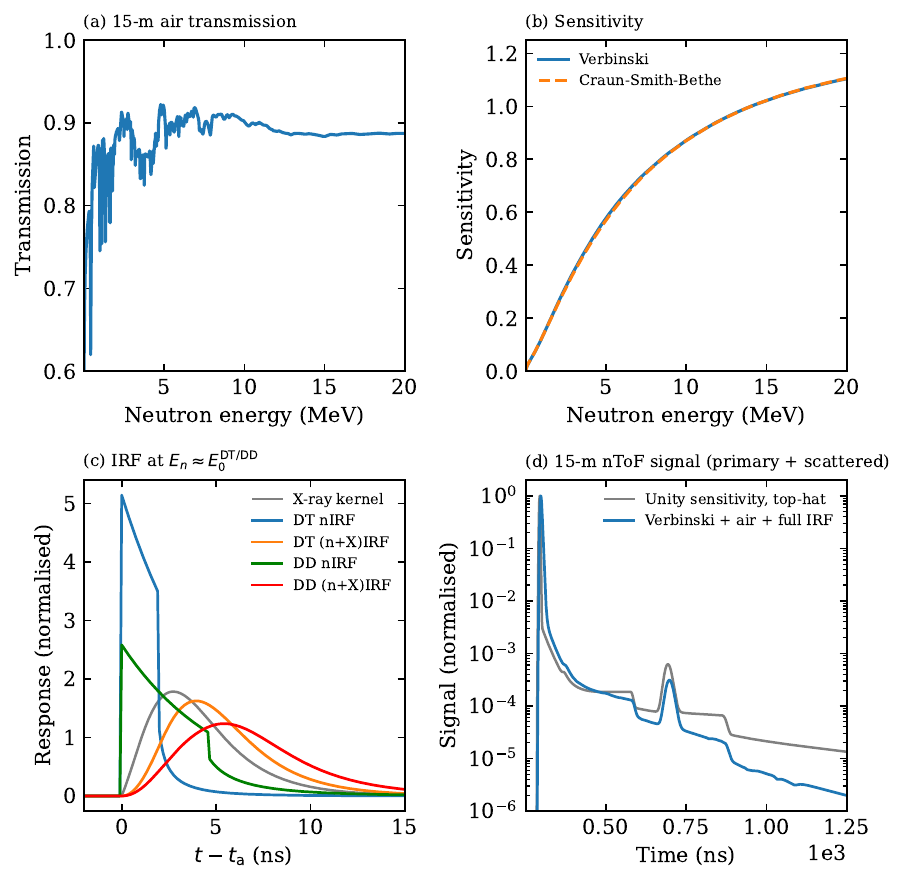}
  \caption{%
    (a) Transmission of 15-m dry air at STP as a function of neutron energy, computed by NeSST from ENDF total cross sections for $^{14}$N, $^{16}$O and $^{40}$Ar.
    (b) Comparison of Verbinski tabulated data with the Craun, Smith and Bethe ($C$ = 0.3, $akB$ = 0.42 MeV, $I$ = 80 eV) scintillator sensitivity model computed using Eq. \ref{eq:sensitivity}.
    (c) Columns of the IRF matrix $\mathbf{R}$ at $E_n = E_0^{DT}$ and $E_n = E_0^{DD}$ for the gated-decaying-Gaussian ($\sigma$ = 2 ns, $\tau$ = 2.38 ns, $t_0$ = 0.66 ns, $\sigma_{gate}$ = 0.8 ns) X-ray kernel only , inverse-Gaussian
    nIRF, and the combined IG~nIRF convolved with the gated-decaying-Gaussian kernel.  The IG~nIRF adds a pronounced late-time tail.
    (d) Synthetic 15-m nToF signal for the DT plasma primary spectrum plus first-scatter continuum at $T_i=5$~keV, $\langle\rho L\rangle=150~\text{mg/cm}^2$: idealised (grey) versus full forward model including Verbinski sensitivity, air attenuation, and composite IRF (blue).  Both curves show the full energy range down to $>$ 1 MeV on a logarithmic scale.
  }
  \label{fig:ntof}
\end{figure}

Combining all terms, Fig.~\ref{fig:ntof} shows the NeSST forward model for a 15-m nToF detector with Verbinski sensitivity, 15-m air attenuation, inverse-Gaussian nIRF base and gated-decaying-Gaussian xIRF kernel.  The composite IRF broadens and adds an asymmetric tail to the DT peak signal relative to the idealised top-hat response.

Listing~\ref{lst:ntof} shows how a full forward model is constructed within NeSST.

\begin{lstlisting}[caption={Construction of a full forward model for neutron time of flight calculation},label={lst:ntof}]
import NeSST.time_of_flight as ntof

detector_distance = 15.0 # m

# 15-m air beamline: dry air at STP (1.225 kg/m^3)
# N.B. user must add ENDF files to the NeSST ENDF directory
air_LOS = [
    ntof.LOS_material(
        density=1.225,
        length=detector_distance,
        components=[
            ntof.LOS_material_component(0.7848, 14.0,  "n-007_N_014.endf"),
            ntof.LOS_material_component(0.2105, 16.0,  "n-008_O_016.endf"),
            ntof.LOS_material_component(0.0047, 40.0,  "n-018_Ar_040.endf"),
        ],
    )
]
air_atten = ntof.get_LOS_attenuation(air_LOS)

# Craun-Smith-Bethe scintillator sensitivity
CSB_sens = ntof.get_CraunSmithBethe_NLO(C=0.3, akB=0.42e6, excitation_energy=80.0)

# Combined sensitivity: CSB x air attenuation
def combined_sensitivity(En):
    return CSB_sens(En) * air_atten(En)

# Gated decaying-Gaussian X-ray detector IRF
xray_kernel_fn = ntof.gated_decaying_gaussian_kernel(
    sig=2.0e-9, tau=2.38e-9, shift_t=0.66e-9, sig_turnon=0.8e-9
)

# Combined IRF: IG-nIRF base convolved with gated-decaying-Gaussian kernel
combined_irf_fn = ntof.make_transit_time_IRF(
    thickness=10e-2,
    kernel_fn=xray_kernel_fn,
    base_matrix_fn=ntof.inversegaussian_nIRF,
)

# Create combined detector model
det_comb = ntof.nToF(detector_distance, combined_sensitivity, combined_irf_fn)
# Use det_comb.get_signal(En, dNdE) method to get time of flight signal
\end{lstlisting}

\subsection{Time resolved nToF signals}

\begin{figure}[!h]
  \centering
  \includegraphics[width=0.8\columnwidth]{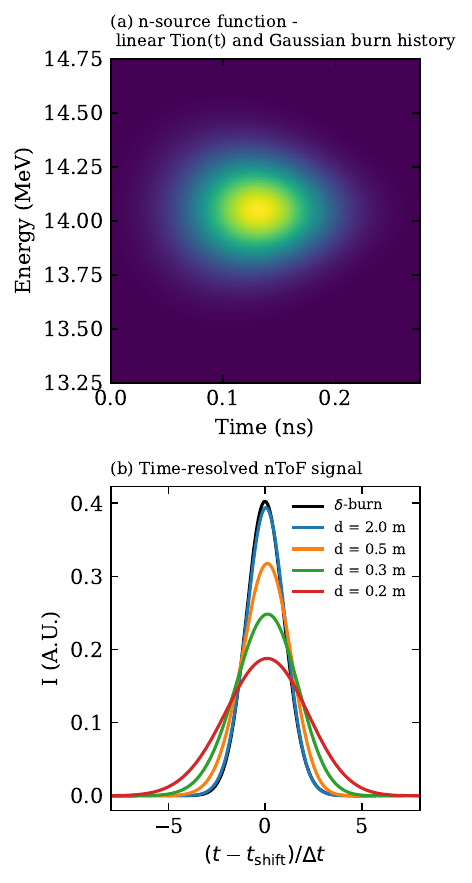}
  \caption{%
    (a) Neutron source function ($d^2N/dEdt$) for a linearly decreasing ion temperature and Gaussian burn history, showing the time evolution of the primary DT spectrum. 
    The primary peak narrows and shifts to lower energy as $T_i$ decreases. The parameters used here are a burn width of 100 ps, central ion temperature of 5 keV, 
    and a linear decrease in ion temperature of 2 keV over the burn width.
    (b) Time-resolved nToF signals at a number of detector distances for the same burn history and ion temperature evolution. Here we use no IRF or sensitivity to highlight temporal birth effects. 
    In black is the measured signal for an instantaneous burn (all neutrons made at same time). Depending on the detector distance, the time-resolved signal varies from the instantaneous burn case, with the most significant differences at short distances. 
    Note the time axis has been scaled to ease comparison between the different distances. The temporal shift combines the mean energy flight time and burn centroid. The temporal width includes only the width from velocity dispersion ($\Delta t = \frac{\sigma_E}{dE/dt}$).
    }
  \label{fig:ntof_TR}
\end{figure}

NeSST also supports time-resolved nToF signals, which can be used to model the effects of finite burn duration and time-evolving plasma conditions on the measured time of flight signals. For current nToF diagnostics, the combination of burn duration and velocity dispersion is usually minimised by placing the detector sufficiently close (e.g. NTD\cite{stoeckl2016}) or sufficiently far (e.g. NIF nToF suite\cite{Moore2021}). However, the combined effect can be significant for intermediate-distance nToF detectors, where the neutron arrival time is comparable to the burn duration. Literature on this topic\cite{catenacci2020,vlad1984,appelbe2024ToF,moore2022} has shown that time-resolved information on plasma conditions can be extracted from nToF signals at multiple distances.  
So far in this work we have considered neutron spectra in a time independent fashion. To include finite burn duration effects we must include the time shift associated with their emission time. The resultant time of flight signal is an effective convolution over emission time:
\begin{align}
  I(t_d) &\propto \int dt_e \int dt_a \ \frac{d^2N}{dE_ndt_e}(E_n, t_e) \ \cdot S(E_n) \nonumber \\
  &\cdot \mathcal{T}(E_n) \cdot \left|\frac{dE_n}{dt_a}\right| \cdot R(E_n, t_d - t_a(E_n) - t_e) 
\end{align}
Where $t_e$ is the neutron emission time, $t_a$ is the neutron arrival time (for emission at t=0), $t_d$ is the detector arrival time, and $\frac{d^2N}{dE_ndt_e}$ is the neutron source function. 

In NeSST, the time convention is that no emission occurs before $t_e < 0$, such that nothing can arrive at detectors before a normalised arrival time $\tau$ of 1 (i.e. before light can reach the detector).

Figure~\ref{fig:ntof_TR} shows the effect on the time-resolved nToF signals of a burn duration of 100 ps with a linearly decreasing ion temperature with a central ion temperature of 5 keV. These finite burn duration effects can be included in the forward model and used to fit time-resolved plasma conditions, including sensitivity and IRF effects, however, this is outside the scope of this work.

\section{\label{sec:fitting}Synthetic Forward-Model Fitting Demonstration}

A key use case of NeSST is generating synthetic nToF signals for comparison with experiment and for fitting plasma parameters.  Here, we demonstrate this fitting capability using a self-consistency test: a full synthetic nToF signal is constructed with the NeSST forward model, realistic Gaussian noise is added, an exponential background is added and the input parameters are recovered via $\chi^2$ minimisation.  This will illustrate the capabilities of NeSST in nToF signal fitting, although application to real experimental data requires instrument-specific IRF calibration, line-of-sight-specific sensitivity and attenuation, and a propagation of calibration uncertainties.

\subsection{Synthetic data construction}

The forward model combines: (i) the DT, DD and TT primary spectra plus first-scatter continuum (including ion velocity effects) evaluated on a common energy grid; (ii) DT and DD peak energies shifted by a hotspot velocity $v_\mathrm{hs}$ via the relativistic velocity-addition formula; (iii) conversion to a time-domain signal through the full nToF forward model of Eq.~(\ref{eq:ntof}) with Craun, Smith \& Bethe (CSB) sensitivity, 15-m air attenuation, and the composite IRF from Sec.~\ref{sec:ntof}.

The measured signal includes a slow empirical background starting at the DT peak arrival time $t_0 = d/v_n(E_0^{DT})$:
\begin{equation}
  I_\mathrm{total}(t) =
    Y_n\,I_\mathrm{signal}(t;\, T_i, \langle\rho L\rangle, v_\mathrm{hs}, \bar{v}, \Delta_v)
    + A_\mathrm{bg}\,e^{-(t-t_0)/\tau_\mathrm{bg}},
\label{eq:bkg}
\end{equation}
where $\tau_\mathrm{bg} = 500$~ns is fixed and $A_\mathrm{bg}$ is a free amplitude.  White Gaussian noise at 3\% of the local signal level is added to obtain the synthetic ``data''. 

\subsection{Parameter recovery}

\begin{figure}[h]
  \centering
  \includegraphics[width=\columnwidth]{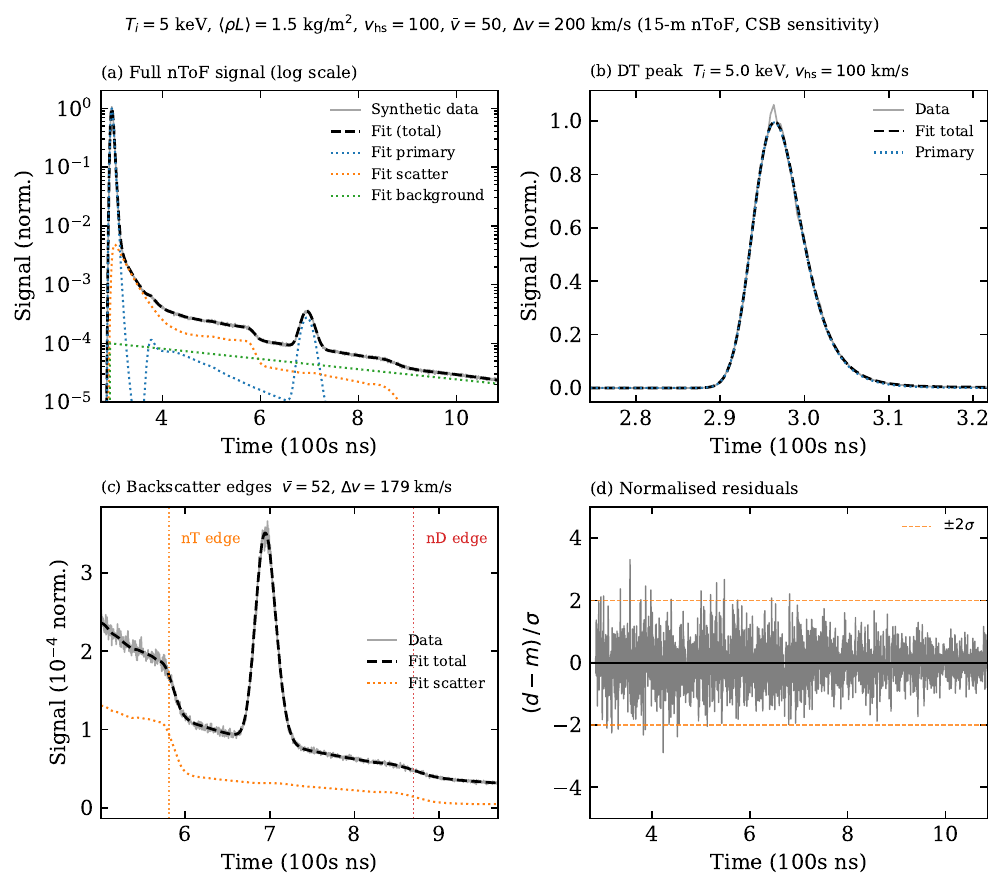}
  \caption{%
    Synthetic code demonstration: forward-model self-consistency test. (a) Synthetic ``data'' (grey) generated with the full NeSST forward model (CSB sensitivity, 15-m air, composite IRF) at $T_i=5$~keV, $\langle\rho L\rangle=1.5$~kg/m$^2$ ($150~\text{mg/cm}^2$), $v_\mathrm{hs}=100$~km/s, $\bar{v}=50$~km/s and $\Delta_v=$200 km/s plus 3\% Gaussian noise and a slow exponential background.  The best-fit
    signal plus background (black dashed), primary component (blue dotted), scatter component (orange dotted) and background component (green dotted) are
    overlaid.  
    (b) Magnified view of the fit to the DT peak. (c) Magnified view of the fit about the backscatter edges and DD peak. (d) Residuals between total fit and synthetic data.
  }
  \label{fig:fitting}
\end{figure}

The free parameters are $T_i$, $\langle\rho L\rangle$, $v_\mathrm{hs}$, edge parameters ($\bar{v}$ and $\Delta_v$), the yield scale $Y_n$, and the background amplitude $A_\mathrm{bg}$.  These are recovered by Powell minimisation of the $\chi^2$ between the synthetic data and forward model.  Figure~\ref{fig:fitting} shows the result for $T_i=5$~keV, $\langle\rho L\rangle=1.5$~kg/m$^2$, $v_\mathrm{hs}=100$~km/s, $\bar{v}=50$~km/s and $\Delta_v=$200 km/s. As expected, the fitted parameters agree with the true values to high accuracy (the edge parameters showing the largest error) and the normalised residuals are consistent with the imposed noise level. We note that in experiment the full spectrum is not typically accessible from a single detector due to required dynamic range, instead multiple detector signals are fit, either independently or combined.

\subsection{Detector model bias}

In addition to synthetic data generation, the NeSST parameterised detector models allow exploration of the sensitivity of fit parameters to detector physics uncertainty. One can perturb the detector model parameters when creating the synthetic data and then fit with the nominal detector model to infer the bias introduced. Here we select a subset of sensitivity and IRF parameters to perturb and look at the effect on the inference of key ICF diagnostics, namely $T_i$, $\langle\rho L\rangle$ and $v_\mathrm{hs}$.

\begin{figure}[h]
  \centering
  \includegraphics[width=\columnwidth]{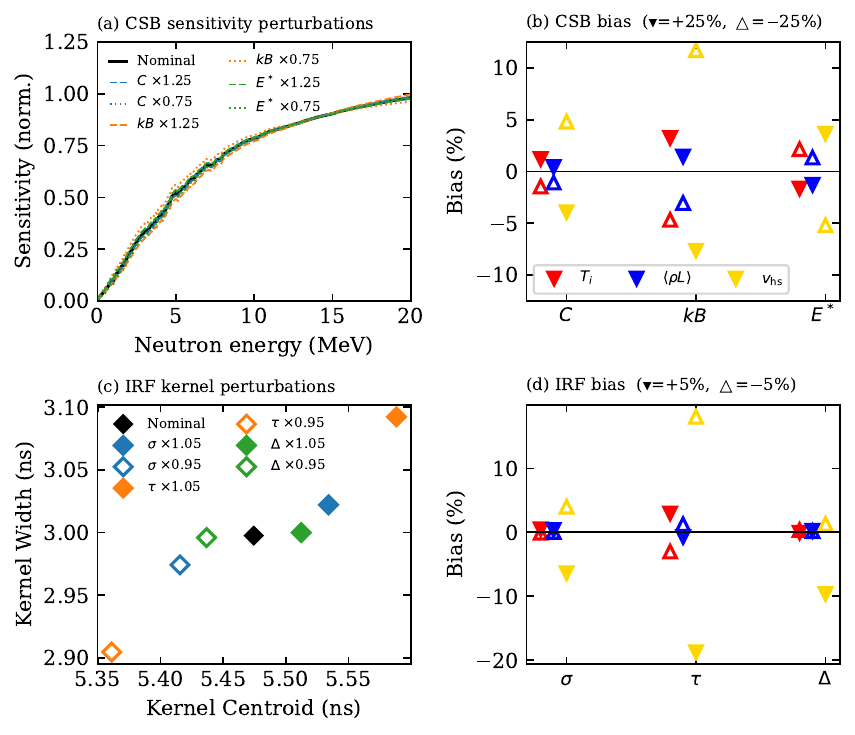}
  \caption{%
    (a) Craun, Smith \& Bethe sensitivity models including perturbations to nominal parameter values. Note that 15m of air is included but not varied. (b) Effect of sensitivity parameter variations on fit values of $T_i$, $\langle\rho L\rangle$ and $v_\mathrm{hs}$. (c) Variations in composite IRF centroid and width (temporal standard deviation) from perturbation in model parameters (note that $\Delta$ is scintillator thickness). (d) Effect of IRF parameter variations on fit values of $T_i$, $\langle\rho L\rangle$ and $v_\mathrm{hs}$.
  }
  \label{fig:bias}
\end{figure}

As shown in Figure~\ref{fig:bias}, significant bias can be introduced through detector model uncertainty. It should be noted that the inferred parameters ($T_i$, $\langle\rho L\rangle$ and $v_\mathrm{hs}$) depend on different features of the signal. These are relative timing ($T_i$ and peak width), relative amplitude ($\langle\rho L\rangle$ and component separation) and absolute timing ($v_\mathrm{hs}$ and peak location). The calculated biases are consistent with these feature differences, for example hotspot velocity inference suffers when any perturbation causes a shift in the centroid of the IRF or detected peak. NeSST enables this type of sensitivity study and can therefore identify which detector model uncertainties must be minimised to enable accurate inference of physical parameters.

\section{\label{sec:implementation}Software Implementation}

\subsection{Package structure}

NeSST is structured as a Python package (\texttt{src/NeSST/}) with the
following modules:
\begin{description}
  \item[\texttt{core.py}] Top-level API functions for primary spectra,
    scattering, transmission, fitting helpers, and yield calculations.
  \item[\texttt{spectral\_model.py}] The \texttt{material\_data} class
    that owns cross section data, energy grids, and scattering matrices
    for each species.
  \item[\texttt{collisions.py}] Relativistic and classical scattering
    kinematics (Lorentz-invariant formulation).
  \item[\texttt{cross\_sections.py}] Differential cross section
    evaluation from Legendre coefficients or tabulated data.
  \item[\texttt{endf\_interface.py}] Reading and parsing ENDF-format
    nuclear data from bundled JSON configuration files.
  \item[\texttt{time\_of\_flight.py}] nToF signal synthesis, IRF
    construction, sensitivity models and LOS attenuation.
  \item[\texttt{dress\_interface.py}] Wrapper for pyDRESS.
  \item[\texttt{constants.py}] Physical constants from CODATA~2018
    via \texttt{scipy.constants}.
\end{description}

\subsection{Computational design}

The scattering matrices representing $(d^2\sigma/dEd\mu_s)$ are evaluated on regular grids and computed once on initialisation and cached.  Scattered spectra at arbitrary areal density and asymmetry are then obtained by matrix--vector multiplication and quadrature, making individual spectrum evaluations fast (typically of order 100s milliseconds to seconds on a modern workstation).  The ion-kinematics kernel adds an additional dimension ($v_i$) to the cached array; scattered spectra at arbitrary $(\bar{v}, \Delta v)$ are obtained by quadrature over this pre-computed table.

\subsection{Dependencies and installation}

NeSST depends on \texttt{NumPy}~\cite{numpy}, \texttt{SciPy}~\cite{scipy}, \texttt{endf-python}~\cite{endf-python} and \texttt{matplotlib}~\cite{matplotlib} as runtime dependencies, all
of which (bar endf-python) are standard in scientific Python distributions.  The package
is installable from the Python Package Index:
\begin{lstlisting}
pip install NeSST
\end{lstlisting}
A development install from the git repository is also supported:
\begin{lstlisting}
git clone https://github.com/aidancrilly/NeSST.git
cd NeSST && pip install -e .
\end{lstlisting}
Documentation is hosted at \url{https://nesst.readthedocs.io}. A selection of automated tests are implemented as part of continuous integration.

\section{\label{sec:summary}Summary}

We have presented NeSST, an open-source Python package for computing
primary and singly scattered ICF neutron spectra and for constructing
synthetic neutron time-of-flight diagnostics.  The key capabilities
are:

\begin{enumerate}
  \item \textbf{Primary spectra}: DT and DD spectral shapes and moments using Ballabio fits, with DRESS Monte Carlo as a reference.  Multiple TT spectral models based on the Brune\cite{brune2015} R-matrix calculation and experimental measurements\cite{GatuJohnson2018,eriksson2024}, all broadened by the Appelbe\cite{appelbe2016TT} Doppler broadening kernel. Yield ratios are calculated using available reactivity models. Relativistic velocity-addition correction available to model hotspot flow.
  \item \textbf{Single-scatter model}: Elastic ($n$D, $n$T) and inelastic [D$(n,2n)$p, T$(n,2n)$D] processes treated via bundled ENDF\cite{ENDF} cross sections with relativistic kinematics.  Additional ablator materials (e.g.\ $^{12}$C, $^9$Be) supported through a general ENDF interface. Approximations for higher $\langle\rho L\rangle$, including double scatter and transmission, are available.
  \item \textbf{Areal density asymmetries}: Full Legendre mode expansion of the neutron-averaged areal density, enabling the spectral effect of implosion non-uniformities to be computed and compared with experimental data.
  \item \textbf{Backscatter edge and ion kinematics}: Pre-computed ion-velocity-dependent scattering kernel allows rapid evaluation of the edge shape as a function of dense fuel temperature and bulk flow velocity.
  \item \textbf{Synthetic nToF diagnostics}: Full forward model\cite{hatarik2015,mohamed2020} [Eq.~(\ref{eq:ntof})] with configurable composite neutron and X-ray IRFs, scintillator sensitivity models, and ENDF-based line-of-sight attenuation.
\end{enumerate}

NeSST is intentionally lightweight: it does not model multi-scatter contributions at full fidelity, spatial profiles, or radiation transport, and detector models require instrument-specific calibration before application to real data.  Within the single-scatter approximation, using cached scattering matrices that reduce spectrum evaluation to cheap table reductions, it provides a fast and flexible tool for neutron spectral analysis and the construction of forward models suitable for ICF diagnostic fitting and synthetic diagnostic generation. For these use cases, users should be aware of the following limitations:

\begin{description}
  \item[Single-scatter approximation] The primary spectral model
    and all singly scattered components are based on the formalism and assumptions
    of Refs.~\cite{Crilly2020,Crilly2021}.  The
    approximate second-scatter and transmission terms are limited in applicability and are not a substitute for proper neutron transport when required.  At $\langle\rho L\rangle\gtrsim 200~\text{mg/cm}^2$ multi-scatter contributions become significant and full transport codes (e.g.\ Minotaur\cite{Crilly2018}, MCNP\cite{mcnp}, OpenMC\cite{openmc}, IRIS3D\cite{Weilacher2018})
    should be used to verify any NeSST results.
  \item[Detector model calibration] The IRF models and sensitivity
    parametrisations provide functional forms that must be calibrated to a specific instrument.  The IRF coefficients provided are fit to unpublished calculations and data and will differ between detectors. The comparisons shown in this
    paper make use of synthetic data.  Application to specific experimental campaigns requires independent validation against measured data and proper treatment of systematic uncertainties such as calibration and backgrounds.
   \item[Nuclear data availability] NeSST relies on available nuclear data to predict the neutron spectra. Some reactions (such as T(n,2n)D) are poorly measured; this represents a systematic uncertainty in NeSST's predictions which must be considered. 
\end{description}

\section*{Acknowledgments}
The author acknowledges financial support from Imperial College London through an Imperial College Research Fellowship grant, and the International Atomic Energy Agency through the AI for Fusion coordinated research project.

The author would also like to thank Dr Owen Mannion and Dr Brian Appelbe for their assistance in developing the theory and methods for neutron spectra within NeSST, Dr Carl Brune for providing his R matrix calculation code, and Dr Jonathan Shimwell for his assistance in improving the software implementation of NeSST.

\bibliography{nesst_refs}

\end{document}